# Impact of technological synchronicity on prospects for CETI

Marko Horvat[1], Anamari Nakić[1], Ivana Otočan[2]

[1]Faculty of Electrical Engineering and Computing, University of Zagreb
Unska 3, HR-10000 Zagreb, Croatia
[2]Faculty of Natural Sciences and Mathematics, University of Zagreb
Bijenička cesta 30, HR-10000 Zagreb, Croatia
E-Mail: Marko.Horvat2@fer.hr, Anamari.Nakic@fer.hr, Ivana.Otocan@gmail.com

**Abstract.** For over 50 years, astronomers have searched the skies for evidence of electromagnetic signals from extraterrestrial civilizations that have reached or surpassed our level of technological development. Although often overlooked or given as granted, the parallel use of an equivalent communication technology is a necessary prerequisite for establishing contact in both leakage and deliberate messaging strategies. Civilization advancements, especially accelerating change and exponential growth, lessen the perspective for a simultaneous technological status of civilizations thus putting hard constraints on the likelihood of a dialogue. In this paper we consider the mathematical probability of technological synchronicity of our own and a number of other hypothetical extraterrestrial civilizations and explore the most likely scenarios for their concurrency. If SETI projects rely on a fortuitous detection of leaked interstellar signals (so called "eavesdropping") then with minimum prior assumptions $N \geq 138 - 4991$ Earth-like civilizations have to exist at this moment in the Galaxy for the technological usage synchronicity probability $p \geq 0.95$ in the next 20 years. We also show that since the emergence of complex life, coherent with the hypothesis of the Galactic habitable zone, $N \geq 1497$ extraterrestrial civilizations had to be created in the Galaxy in order to achieve the same estimated probability in the technological possession synchronicity which corresponds to the deliberate signaling scenario.

## 1 Introduction

All extraterrestrial (ET) signals from hypothetical civilizations may be categorized in two discrete classes according to their intent to communicate: 1) intentional or deliberate signals which are purposely sent towards Earth (i.e. the receiver); and 2) unintentional or inadvertent signals which are involuntarily and accidentally directed towards the Earth's position. The first class implies intention to communicate, while the latter represents so called leakage or leaked signals analogous to our own radio, TV or radar [1][2][3].

Since the start of Project Ozma in 1960 Search for Extraterrestrial Intelligence (SETI) and Communication with Extraterrestrial Intelligence (CETI) experiments were designed to find microwave radio and optical signals intentionally sent from ET civilizations [4][5]. The wish for interstellar communication was either assumed, or the probability of finding a transmitting civilization was expected to average out in the large dataset involved. Detection and verification of an intermittent, let alone unique, signal is difficult or even impossible.

Very important aspect of SETI searches is *the synchronicity problem* which considers prospects of simultaneous transmissions of ET signals and reception by our receivers. This requirement is embedded in all contemporary SETI programs and a positive outcome is necessary for successful detection of ET messages if they do exist [6]. The synchronicity problem is aligned with the technology used on both sides of the communication channel. If actors at the transmitting and receiving end do not use the same communication method and determining parameters of the method, they cannot detect each other or exchange information regardless of a possible continuous or omnidirectional use of their communication devices.

In this paper we use geometric probability to model prospects of the technological synchronicity between $N \in \mathbb{N} \cup \{0\}$ civilizations which leak signals. Our model leads to following conclusions. If we assume SETI projects have a good chance of making a detection in the next few decades [7][8] then either *i)* our neighborhood is populated with a large number of technology compatible civilizations in the leakage scenario, or this scenario is wrong and *ii)* advanced civilizations intentionally broadcast greeting messages across the Galaxy. As a completely alternative speculation, *iii)* both communication scenarios are wrong (leakage and intentional) and advanced civilizations are in effect frugal emitting extremely low-level or unidirectional signals which may be undetectable by our present of projected near-future receivers [9]. However, neither scenario precludes advanced civilizations from monitoring signals made by our current level of technology.

## 2      Technological synchronicity problem

The technological synchronicity problem is defined as an independent, accidental and simultaneous possession or use of compatible technologies, including those for interstellar communication, among two or more civilizations. From this we can identify two subtypes of the problem. The first type is the possession synchronicity which is defined around possession or knowledge of a technology. The second type is the technological usage synchronicity problem which deals with application and practical use of a particular technology. By definition the usage should include the possession problem: it's not possible to use a technology without first having it and knowing how to apply it. Therefore the period of a technology possession is as long or longer then the period of its usage. This makes the possession synchronicity problem more relaxed than the application, i.e. usage, problem. For example, the possession period of bronze artifacts is ~5300 years but the usage period was just ~2000 years as the metal smelting technology jumped from bronze to iron [10]. On the other hand, radio has been first applied for long-range signaling in 1895 by Guglielmo Marconi who based his work on previous achievements by Nikola Tesla and other researchers like Heinrich Hertz [11][12]. Approximately 120 years later we are still actively using radio which makes its possession and usage periods equal. Although the knowledge of a particular technology can be very long, it may be rendered obsolete relatively quickly because of the discovery of superior technologies. This incessant process is governed by the rate of civilization's technological advancements.

So far persistent technological progress of our civilization is an incontestable fact. The global technological progress may be estimated directly with the rate of inventions or indirectly through representative correlates and dependent values such as the world population growth, food consumption, energy use, but also the number of granted patents, per capita income, etc [13][14][15]. Our rate of development, although often heterogeneous, over sufficiently long periods has so far been non-stagnant and positive. The pool of technologies at the civilization's disposal has been increasing steadily. Furthermore, there are some indications that in the recent times our technological progress is accelerating and growing exponentially [15]. We cannot predict with certainty where this process will finally lead to. The rate of progress might reach a certain threshold (also called technological singularity) after which it will change so profoundly that it cannot be fully comprehended at this point. But also, hypothetically, a different kind of a barrier may exist that prevents, or at least significantly prolongs, development of technologies for interstellar communication better than microwave radio. Both opposing scenarios should be taken into account in modeling the synchronicity problem.

According to the Copernican principle there is nothing special about our civilization or its location in the Galaxy. If we recognize this rule to be correct then our civilization is not a fluke or a statistical outlier and may be used as a statistically representative template in modeling other hypothetical ET civilizations. Therefore, we can assume that ET societies also have significant technological progress, at least over long time scales, and that *the technological synchronicity problem is ubiquitous and inherent to the interstellar communications.*

## 3  Geometric probability model

The synchronization of any two time intervals can be successfully described with mathematical models based on geometric probability. This can be exploited for explicit description of the simultaneous use of compatible technologies for signaling over interstellar distances.

Our geometric probability model describes temporal synchronization between two mutually independent random events: emitting of a signal by the sender S and monitoring for a signal by the receiver R. These events are represented with two random variables: $X_1$ and $X_2$. The first variable $X_1$ represents the moment when S begins signaling and the second variable $X_2$ is the moment when R begins to listen. There are also two intervals ($\tau_1$ and $\tau_2$) which are constants. These intervals are the periods during which S is signaling and R is monitoring for signals, respectively. Therefore, S is signaling during interval $[x_1, x_1+\tau_1]$ and R is monitoring during $[x_2, x_2+\tau_2]$.

The detection *will not take place* if and only if S stops emitting before R starts listening, or if S starts emitting after R stopped listening (Figure 1). These rules can be written as

1. condition "S stops before R":     $x_1 + \tau_1 < x_2$
2. condition "S starts after R":     $x_2 + \tau_2 < x_1$

(1)

In any other case there will be at least some overlapping between these two intervals and the signal from S could be detected. This is exactly the event we are interested, i.e. that represents the technological synchronicity between a single sender – receiver pair. Because we assume that S and R will occur within well defined intervals

Signaling start:   $x_1 \in [0, T - \tau_1]$

Reception start:   $x_2 \in [0, T - \tau_2]$

(2)

Thus, the total sample space is

$$\Omega = [0, T - \tau_1] \times [0, T - \tau_2] \tag{3}$$

The positive synchronization event A between one sender and one receiver is determined by inversing inequalities (1) that define event spaces $G_1$ and $G_2$ through their respective measures of geometric spaces $m(G_1)$ and $m(G_2)$, as well as the measure of the total sample space $m(\Omega)$ (Figure 2). In a plane space which we use here the measure is represented by a two dimensional area which can be derived geometrically. Union ($\cup$) of two or more sets represents their joint sample space, while intersection ($\cap$) is their common area.

Therefore, if $(T - \tau_1 - \tau_2) > 0$

$$P(A) = \frac{m(\Omega) - m(G_1) - m(G_2)}{m(\Omega)} = 1 - \frac{\frac{1}{2}(T - \tau_2 - \tau_1)^2 + \frac{1}{2}(T - \tau_1 - \tau_2)^2}{(T - \tau_1)(T - \tau_2)} \tag{4}$$

$$P(A) = 1 - \frac{(T - \tau_1 - \tau_2)^2}{(T - \tau_1)(T - \tau_2)} \tag{5}$$

Note that $(T - \tau_1 - \tau_2) \leq 0 \Rightarrow P(A) = 1$ because $(G_1 \cup G_2) \cap \Omega = \emptyset$.

This model can be extended so it takes into account $S_1, \ldots, S_N$ devices which emit signals and one receiving device R. This is a more realistic and pragmatically applicable model. We define events $A_i, i = 1, \ldots, N$ where $A_i$ represents device R receiving a signal sent by device $S_i, i = 1, \ldots, N$. The synchronization event A is positive if *at least one* $S_i$ has received a signal from R. The receiver can be synchronized and overlapping with one or more transmitters. Therefore, A is realized if any of the events $A_1, \ldots, A_N$ is also realized: $A = A_1 \cup \ldots \cup A_N$. The probability

P(A) is derived from Sylvester formula which is also sometimes called the inclusion-exclusion principle [16]

$$P(A) = P\left(\bigcup_{i=1}^{N} A_i\right) = \sum_{i=1}^{N} P(A_i) - \sum_{0 \leq i < j \leq N} P(A_i A_j) + \sum_{0 \leq i < j < k \leq N} P(A_i A_j A_k) - \ldots$$
$$+ (-1)^{N+1} P(A_i A_j \ldots A_N) \quad (6)$$

This formula describes a joint intersection of $A_i, i = \{1, \ldots, N\}$ events. If $A_1$ and $A_2$ are two finite sets then $|A_1 \cup A_2| = |A_1| + |A_2| - |A_1 \cap A_2|$, for three finite sets $|A_1 \cup A_2 \cup A_3| = |A_1| + |A_2| + |A_3| - |A_1 \cap A_2| - |A_1 \cap A_3| - |A_2 \cap A_3| + |A_1 \cap A_2 \cap A_3|$, and so on. The annotation $|A_i|$ describes the cardinality of the set $A_i$, i.e. the probability of the event $A_i$. Therefore, Sylvester formula is used to describe the probability of a cumulative synchronization event A which is composed of N singular synchronization events each between a single sender-receiver pair.

Events $A_i$ and $A_j$ are mutually independent, thus $P(A_i A_j) = P(A_i) P(A_j)$ and we can write

$$P(A_i) = 1 - \frac{(T - \tau_1 - \tau_2)^2}{(T - \tau_1)(T - \tau_2)} \Rightarrow P(A_i A_j) = \left[1 - \frac{(T - \tau_1 - \tau_2)^2}{(T - \tau_1)(T - \tau_2)}\right]^2 \quad (7)$$

Furthermore

$$P(A_i A_j A_k) = \left[1 - \frac{(T - \tau_1 - \tau_2)^2}{(T - \tau_1)(T - \tau_2)}\right]^3 \quad (8)$$

Generally, for every $k$ synchronization events, $k = \{1, \ldots, N\}$

$$P(A_{i_1} A_{i_2} \ldots A_{i_k}) = \left[1 - \frac{(T - \tau_1 - \tau_2)^2}{(T - \tau_1)(T - \tau_2)}\right]^k \quad (9)$$

When we include this equation in (6) we get

$$P(A) = N\left[1 - \frac{(T - \tau_1 - \tau_2)^2}{(T - \tau_1)(T - \tau_2)}\right] - \binom{N}{2}\left[1 - \frac{(T - \tau_1 - \tau_2)^2}{(T - \tau_1)(T - \tau_2)}\right]^2 +$$
$$\binom{N}{3}\left[1 - \frac{(T - \tau_1 - \tau_2)^2}{(T - \tau_1)(T - \tau_2)}\right]^3 - \ldots + (-1)^{N+1}\binom{N}{N}\left[1 - \frac{(T - \tau_1 - \tau_2)^2}{(T - \tau_1)(T - \tau_2)}\right]^N \quad (10)$$

For a given $T$, $\tau_1$ and $\tau_2$ we want to assess $N$ so P(A) is larger than a certain threshold (e.g. at least 0.95). The model is used to narrow the least possible number of ET civilizations if the probability of synchronization and its member variables are presupposed *ad hoc*.

Furthermore, since the basic synchronization problem occurs between a single receiver R and a single sender S, and it is divided into two sub-problems (possession and usage) there are $2^2$ discrete combinations which can occur with the sender-receiver pair (Table 1). Both sender and receiver may be using or possessing the same technology, but sender may be possessing and the receiver using it and vice-versa. This is important in developing technology interaction models since the possessing side is intuitively more advanced, but also in developing models

In order to compute the series (10) we developed our own software based on Microsoft .NET 4.0 framework. The application was written in C# object-oriented computer language using MAPM library [17] to computationally express very large numbers with arbitrary precision. Several heuristic optimizations and memory vs. speed trade-offs were necessary to limit the duration of the algorithm's execution.

## 4      Applying the figures

By definition, during the technology possession period a civilization has a particular technology at its disposal but does not necessary have to use it. The possession period can be defined as $\tau_{possess} = t_{possess}^{end} - t_{possess}^{begin}$, where $t_{possess}^{begin}$ and $t_{possess}^{end}$ are the instants of technology acquisition and discontinuation of possession, respectively. We will assume that civilizations have mutually independent technical development which makes the acquisition equivalent to the invention of a technology. In this model civilizations develop technologies on their own without information about technological state of others. Also, barring adverse social events such a retrograde society development, we can assume that $t_{possess}^{end}$ occurs with the death of a civilization.

If a civilization possesses a particular piece of technology during nearly its entire lifetime $t_{possess}^{end} \gg t_{possess}^{begin}$ we can say that the technology in indigenous to that civilization. As could be intuitively expected, the synchronicity with older and more advanced civilizations depends only on the Mankind's technological development – they have the technology already and we have to catch up. The same trend can be observed even with our own microwave radio technology. The technology acquisition moment of radio is $t_{possess}^{begin} = 2 \cdot 10^5 - 1.2 \cdot 10^2 \approx 2 \cdot 10^5$. Using Gott's *delta t* argument [18] we can say with five sigma confidence that the expected lifetime of the Mankind is $\sim 8 \cdot 10^6$. Therefore, the possession period for radio technology is $\tau_{possess} = 8 \cdot 10^6 - 2 \cdot 10^5 = 7.8 \cdot 10^6 \approx 8 \cdot 10^6$.

In this analysis Gott's argument is very helpful, because it puts minimal prior requirements on the temporal variables in the model, but nevertheless it should be used carefully in view of its limitations. The argument rests on the assumptions that a population has uniform distribution and that a choice of a specific individual within the population is random. The chosen individuals ($\tau_{possess}$ and $\tau_{use}$) do not in any way provide information about the population size (the maximum values of $\tau_{possess}$ and $\tau_{use}$). Since its publication the argument has provoked numerous critiques – for example that the expected size of the population has to be infinite (see also [19][20][21]) However, if taken together with the mediocrity principle the delta *t* argument is useful since it yields a statistically significant estimation and does not require extensive research in exobiology with actual determination of as yet unknown variables $\tau_{possess}$ and $\tau_{use}$. It should also be noted that Gott's argument will give an upper bound in the technology usage and the highest probability of the synchronization. But if a civilization follows an exponential technological growth usage of individual technologies will be shortened, which means that the *delta t* argument actually favors the technological synchronization.

In modeling the possession synchronization $T$ represents a larger period within which civilizations can be born. At the most $T$ is equal to the age of the Milky way galaxy ($\sim 13.2 \cdot 10^9$), but in line with the concept of the Galactic habitable zone (GHZ) [22] $T = 4 \cdot 10^9 - 8 \cdot 10^9$.

In the first analysis we assess the possession synchronization probability with exactly one ET civilization using equation (5) and inserting $\tau_1 = \tau_2 = \tau_{possess}$. The chances of synchronization are very slim: $T = 4 \cdot 10^9 \rightarrow p = 0.004$, $T = 8 \cdot 10^9 \rightarrow p = 0.002$ and $T = 13.2 \cdot 10^9 \rightarrow p = 0.0012$. Therefore, even if the lifespan of two Galactic civilizations is on average $8 \cdot 10^6$ years, there is just $\sim 0.12 - 0.4\%$ probability of their concurrent possession of the same technological advancements. The likelihood of the synchronization can be increased if civilizations live longer or if the age distribution for the complex life that may inhabit our Galaxy is smaller, i.e. if $\tau_{possess}$ is larger or $T$ smaller, respectively.

In the second analysis of the technological possession synchronization by using equation (10) we will determine the least number of ET civilizations necessary for $p \geq 0.95$ with different representative values of $T$, $\tau_1$ and $\tau_2$. The results are shown in Figure 3. We see that since the prerequisites for the complex-life have been met, only $N = 747 - 1496$ civilizations similar to our own ($\tau_1 = \tau_2 = 10^8$) should have been formed in the whole Galaxy to get $p \geq 0.95$ in the technological possession synchronicity. In the absolute worst case $T = 13.2 \cdot 10^9$ no more than $N = 2470$ civilizations are necessary.

The simultaneous use of a technology is a very similar but more bounded problem then the possession. It is also more important for SETI since we can expect to receive a signal only from those civilizations that actively use compatible set of communication technologies.

Applying the *delta t* argument we may assume with 95% confidence that our civilization will use microwave radio technology for $\tau_{use} = 123 - 4800$ years in total. For the laser and fiber optical communications which were invented, although not immediately used, ~50 years ago there is also 95% confidence that they will be used for up to $\tau_{use} \approx 2000$ years.

In the third analysis (Figure 4) we are comparing Earth-like civilizations which simultaneously possess and use microwave radio technologies $\tau_1 = 10^8$, $\tau_2 = \tau_{use} = 4.3 \cdot 10^3$ with $T = \{4 \cdot 10^9, 8 \cdot 10^9, 13.2 \cdot 10^9\}$. This represents the human race during the entire projected usage of radio with civilizations that possess and possibly use it until their death. The number of such ET societies needed to achieve very high probability of synchronization is $N = 1496 - 2993$ and $N = 4939$ in the worst case.

More interesting is the fourth analysis (Figure 5) where we are investigating the likelihood of synchronization with advanced Earth-like civilizations ($\tau_1 = 10^8$) in the next couple of decades ($\tau_2 = 20$). Results of the previous and this analysis are very similar because in both cases $\tau_1 - \tau_2 \approx \tau_1$. Only $N = 1497$ civilizations had to be created in the Galaxy during the last $T = 4 \cdot 10^9$ years to get over 95% probability that at least one of them will co-exist with us in the next 20 years.

The last three models presuppose that our and other hypothetical civilizations were created approximately at the same time $T = 2 \cdot 10^5$ years ago. In Figure 6 the usage of microwave radio is exclusively compared $\tau_1 = \tau_2 = \tau_{use} = 4.3 \cdot 10^3$. We see that only $N = 68$ is enough for the targeted probability $p \geq 0.95$. This is remarkable because only a handful of similar civilizations in the Galaxy are enough to get an opportunity to engage in the interstellar messaging with microwave radio.

The two models that are the most directly useful to CETI calculate the probability of synchronization $p$ with an Earth-like civilization actively using microwave radio in the next 20 years. In Figure 6 ET societies are using radio to the maximum $\tau_1 = \tau_{use} = 4.3 \cdot 10^3$, $\tau_2 = 20$ and just $N = 138$ ET civilizations in the Galaxy are enough for $p \geq 0.95$ in this case. Finally, the last model (Figure 7) simulates rapid progress and quick rejection of radio technology $\tau_1 = 100$, $\tau_2 = 20$. Here as many as $N = 4991$ are necessary, which indicates a strong dependence between $N$ and $\tau_1$ confirming it is difficult to synchronize with a civilization that is radio-loud only for a short period.

## 5     Discussion

The technological synchronicity only provides an initial opportunity for CETI. It does not guarantee that interstellar messaging will take place or that the signals will be detected. However, on the basis of the results obtained from our model we can conclude that the technological synchronicity is not a rare event if many ET civilizations do exist in the Galaxy, or they have very long lifespan and intentionally use technologies which are compatible with our stage of development. Another circumstance that might support synchronization is the bounded technological development curve.

In the definition of the geometric probability model (Section 3) we assumed that sender S and receiver R are independent of each other. But by the very nature of the GHZ theory creation of civilizations is clustered and not completely uniform throughout the Galaxy (spatially) or during the Galaxy's development timeline (temporally). However, this does not interfere with the initial postulate of S-R independence since civilizations could be created at the same (cosmological) instant but have different progress curves. The only requirement on the technological synchronization is that S and R invent their technologies independently of each other, or in other words, that they do not come into any sort of contact or exchange information before their technological synchronization is asserted. If these conditions are met then the developed model is valid. Indeed, the mathematical model in this paper can be used to assert the probability of technological synchronization within any time frame – in the last three developed models in the section 4 we assumed concurrently created and young civilizations, while other model's instances have uniform temporal distribution within several billion years.

As has been suggested [23] spatial and temporal coincidence of civilizations' inception and development paths is important for their synchronization and possible contact. These factors certainly confine the simpler model adopted in this paper and act as filters in the probability function. Intuitively, one should expects to see "bursts" of higher synchronization probability within certain time periods and Galaxy regions suitable for generation of life and development of advanced civilizations. However, this argument should be correlated with the GHZ hypothesis or, even more generally, the evolution of habitable planets.

On the other point, the finite speed of light and large distances involved in CETI imply a possibility that R might receive a microwave radio message from S when actually the sending civilization has broke the established synchronization. This should be noted when considering the model presented in the paper, but ultimately it does affect the estimated number of civilizations $N$ necessary for the statistically significant probability of technological synchronization. The finite speed of transmitting messages entails a mixed clique of technologically synchronized ET civilizations where at a specific point in time some either do not exists any more, have superior set of technologies or even receded to a more primitive technological status because of a cosmic cataclysm or some destructive conflict. The number of civilizations $N$ incorporates all these civilizations.

If we assume that a microwave radio contact (or optical, neutrino, etc.) with ET civilizations will be established in the next 20 or so years we should adopt at least one of the following two hypotheses: 1) either there are plenty of relatively juvenile civilizations randomly distributed in the Galaxy at or near our own stage of the technological development and their leaked microwave radio emissions are detectable from Earth, or 2) other civilizations, although possibly more advanced than us, are actively seeking contact and flooding space with intentional messages, i.e. "greetings". In the latter case these civilizations may target specific planets which are good candidates for the development of the complex life [24]. The first hypothesis implies three likely possibilities: 1a) our civilization and a majority of others in the Galaxy were created approximately at the same time and have similar development curves, 2a) civilizations have a short life-span but the sentient life is not sparse and many new radio-capable civilizations are created regularly, and finally, 3a) there is a kind of technological barrier beyond microwave-radio and optical technology which does not allow a more efficient or faster-than-light messaging across interstellar distances, in effect stalling the development of communication technologies at or near our current level of progress.

Civilizations that only inadvertently leak signals should be closer to our stage of development then more advanced civilizations that have better technologies and can muster enough resources to continuously send strong intentional signals. In the cosmological timescales we are at the beginning of the interstellar communication age so there is a lesser chance that we can find another civilization identical to our own. Indeed, if we currently co-exist with other ET civilizations it is likely that they are either at a lower or higher development stage then us.

Furthermore, the minimum required number of advanced civilizations engaged in the deliberate messaging scenario is relatively low but could actually be much higher than what can be deduced strictly mathematically since the will to actively seek contact, as in the Messaging Extraterrestrial Intelligence (METI) effort [25], is optional. This is why the actual number of civilizations in the Galaxy could actually be much higher if SETI does discover an intentional signal. On the other hand, the microwave radio leakage is by definition involuntary and unintentional. Therefore, the number of ET civilizations could be more accurately predicted along this scenario.

We can also deduce that the cumulative duration of SETI projects is insignificant when compared to the larger timescales related to progress and development of civilizations. Consequentially, if we are not technologically synchronized at a particular moment in time with another ET civilization then several decades more or less of SETI effort will not make a statistically significant change. In this respect we can say that SETI implements a static census of the technologically comparable ET civilizations and is oblivious to their dynamical changes.

Finally, it is obvious that the technological synchronicity is a reflective property between two civilizations. Earth is not only a receiver but also a transmitter. Since in the discussion we assumed that there are $N \gg 0$ civilizations in the Galaxy, we can also presume that other ET civilizations do potentially exist that could be using microwave radio communication technologies for detection of Earth's signals. As was mentioned before, if $\tau_{possess} > 0$, $\tau_{use} > 0$ (they can use radio) and $\tau_{possess} \gg \tau_{use}$ is expected statistically, then such civilizations will likely be more advanced than us.

## 6      Conclusion

We have seen that the technological synchronicity is a multifaceted problem but it is also important to SETI as a prerequisite for detection of ET signals. We have divided the synchronicity problem in two classes: a more general class of the synchronous possession of technologies, and – the class that is more interesting to the SETI/CETI effort – the synchronous usage of technologies. These classes of problems were analyzed with our mathematical model based on geometric probability with the goal of $\geq 95\%$ probability of ET signal reception in the next couple of decades.

In the context of positive detection along the leakage scenario we should expect to find a large number of technologically similar civilizations and the intentional signaling allows for a much smaller number of more advanced civilizations. This would make detection of a deliberate and information-rich message more likely but only if ET civilizations are indeed using this resources for this effort.

Furthermore, we saw that only a handful of ET civilizations ($N \geq 138 - 4991$) living in the Galaxy right now are enough to get a fair opportunity of an accidental technological synchronization, yet so far no interstellar signal has been detected. This represents a kind of conundrum and only deepens the Fermi paradox. It can be explained by a small real number of ET civilizations, distance from Earth or their radio quietness which may come from frugality or prudence.

Apart from detecting interstellar signals along either scenario, equally intriguing outcome of the contemporary SETI effort would be a null result. But even in the case of negative detection a well-designed SETI experiment would be an important and scientifically valid because it bounds or excludes different hypotheses on the sentient life and CETI in general.

Pragmatically speaking, the synchronicity problem will certainly play an important role in determining the chances of contact by modern large radio-telescopes such as Low-Frequency Array (LOFAR), Mileura Widefield Array Low Frequency Demonstrator (MWA LFD) and Square Kilometer Array (SKA) [26][27][28][29][30]. The possession and usage of the same technologies determines the actual number of potential CETI targets.

## Acknowledgments

The authors wish to thank their colleagues at University of Zagreb and Croatian Astronomical Society for their cooperation and also an anonymous referee whose comments have greatly improved this paper.

## 7   Tables

| No. | Sender | Receiver |
|---|---|---|
| 1. | $\tau_{possess}$ | $\tau_{possess}$ |
| 2. | $\tau_{possess}$ | $\tau_{use}$ |
| 3. | $\tau_{use}$ | $\tau_{possess}$ |
| 4. | $\tau_{use}$ | $\tau_{use}$ |

Table 1 – Possible combinations in technology synchronization for a single sender-receiver pair. Two actors and two discrete technology synchronization types give $2^2$ different combinations which has affects on the probability of their accidental concurrency.

## 8   Figures

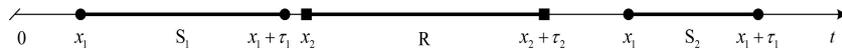

Figure 1 – Absence of detection between a signal receiver R and two emitters $S_1$ and $S_2$. Any other configuration would lead to synchronization between R and $S_i$.

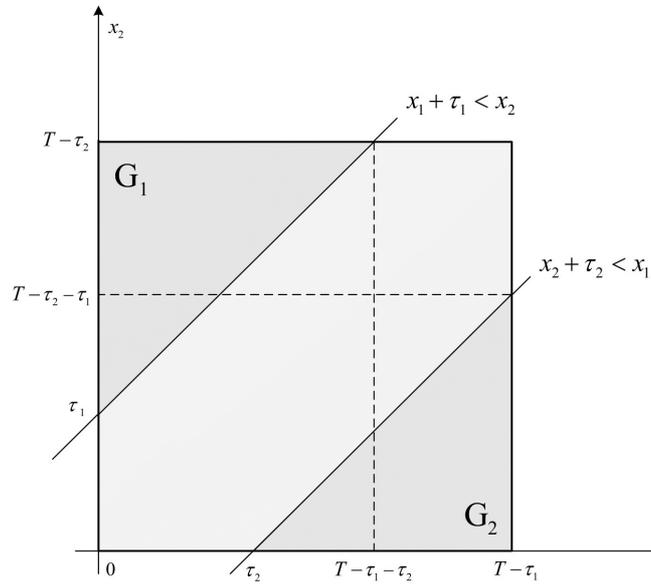

Figure 2 – Probability space of random variables signaling start $X_1$ and reception start $X_2$

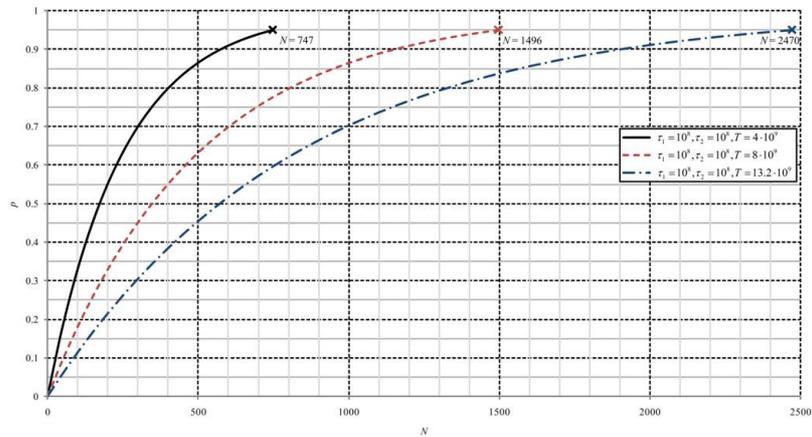

Figure 3 – The least number $N$ of Earth-like civilizations which have to coexist with the human race to achieve at least one synchronicity in any technology ($\tau_1 = \tau_2 = 8 \cdot 10^6$, $T = \{4 \cdot 10^9, 8 \cdot 10^9, 13.2 \cdot 10^9\}$).

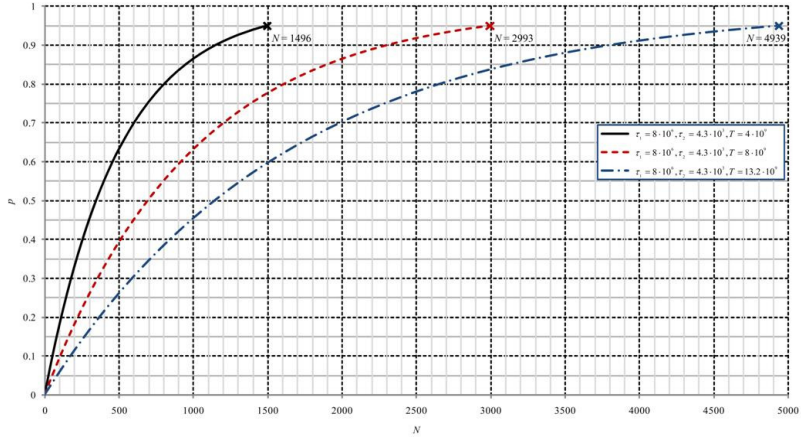

Figure 4 – The least number $N$ of Earth-like civilizations necessary to achieve at least one technological synchronicity $\tau_1 = 8 \cdot 10^6$ during our expected usage of microwave radio $\tau_2 = 4.3 \cdot 10^3$ with $T = \{4 \cdot 10^9, 8 \cdot 10^9, 13.2 \cdot 10^9\}$.

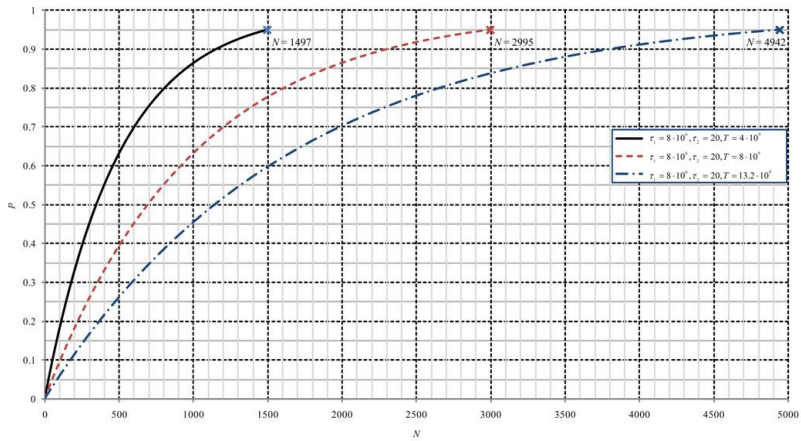

Figure 5 – The least number $N$ of Earth-like civilizations necessary to achieve at least one technological synchronicity $\tau_1 = 8 \cdot 10^6$ in the next $\tau_2 = 20$ years with $T = \{4 \cdot 10^9, 8 \cdot 10^9, 13.2 \cdot 10^9\}$.

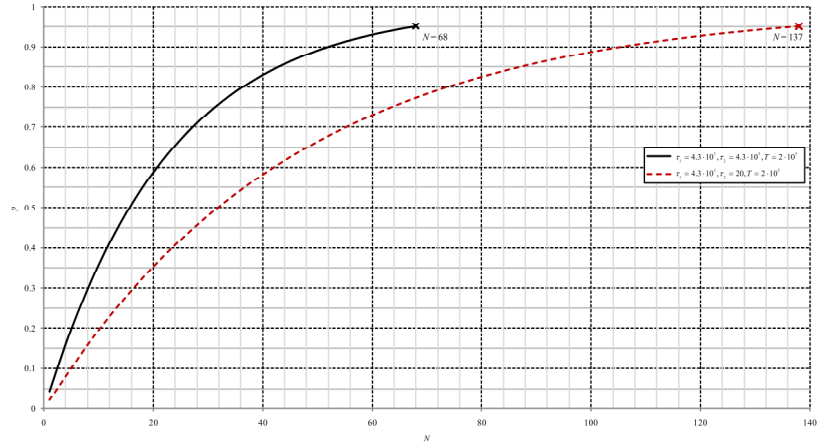

Figure 6 – The least number $N$ of ET civilizations born in the last $T = 2 \cdot 10^5$ years which are required for at least one technological usage synchronicity $\tau_1 = 4.3 \cdot 10^3$ during our expected usage of microwave radio $\tau_2 = 4.3 \cdot 10^3$ and in the next $\tau_2 = 20$ years.

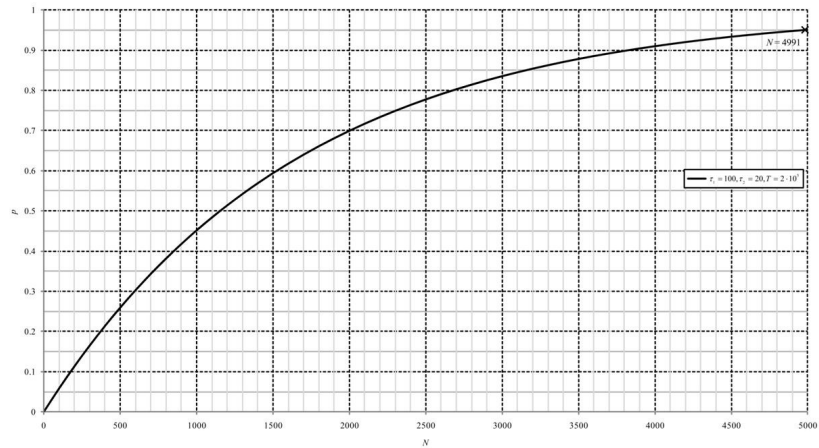

Figure 7 – The least number $N$ of ET civilizations born in the last $T = 2 \cdot 10^5$ years with short technology usage period $\tau_1 = 100$ (exponential development) required for at least one synchronicity in the next $\tau_2 = 20$ years.